\documentclass[aps,groupedaddress,showpacs]{revtex4}
\usepackage{graphicx}
\begin{document}
%%%%%%%%%%%%%%%%%%%%%%%%%%%%%%%%%%%%%%%%%%%%%%%%%%%%%%%%%%%%%%%%%%%%%%%%%%%%%
% You should use BibTeX and revtex.bst for references
%\bibliographystyle{apsrev}
%%%%%%%%%%%%%%%%%%%%%%%%%%%%%%%%%%%%%%%%%%%%%%%%%%%%%%%%%%%%%%%%%%%%%%%%%%%%%
% marks overfull lines with blackboxes
%\draft - no longer supported, use the 'draft' option instead
%\preprint{}
\title{NMR relaxation of quantum spin chains in magnetic fields}
% Optional argument for running titles on pages
%\title[]{}
\author{Takahumi Suzuki}
%\thanks{}
\altaffiliation[Present Address: ]{ISSP, University of Tokyo, Kashiwanoha 5-1-5, Kashiwa, Chiba 277-8581, Japan}
\affiliation{Department of Applied Physics, Osaka University, Suita, Osaka 565-0871, Japan}
\author{Sei-ichiro Suga}
\affiliation{Department of Applied Physics, Osaka University, Suita, Osaka 565-0871, Japan}
\date{\today}
%%%%%%%%%%%%%%%%%%%%%%%%%%%%%%%%%%%%%%%%%%%%%%%%%%%%%%%%%%%%%%%%%%%%%%%%%%%%%
%                         ABSTRACT                                          %
%%%%%%%%%%%%%%%%%%%%%%%%%%%%%%%%%%%%%%%%%%%%%%%%%%%%%%%%%%%%%%%%%%%%%%%%%%%%%
\begin{abstract}
We investigate NMR relaxation rates $1/T_1$ of quantum spin chains in magnetic fields. Universal properties for the divergence behavior of $1/T_1$ are obtained  in the Tomonaga-Luttinger-liquid state. The results are discussed in comparison with experimental results.
\end{abstract}
%%%%%%%%%%%%%%%%%%%%%%%%%%%%%%%%%%%%%%%%%%%%%%%%%%%%%%%%%%%%%%%%%%%%%%%%%%%%%
%\maketitle must follow title, authors, abstract and \pacs
\maketitle
%\section{}
%\label{}
%\subsection{}
%\subsubsection{}
%%%%%%%%%%%%%%%%%%%%%%%%%%%%%%%%%%%%%%%%%%%%%%%%%%%%%%%%%%%%%%%%%%%%%%%%%%%%%
%                        MAIN TEXT                                          %
%%%%%%%%%%%%%%%%%%%%%%%%%%%%%%%%%%%%%%%%%%%%%%%%%%%%%%%%%%%%%%%%%%%%%%%%%%%%%
\section{Introduction}
%%%%%%%%%%%%%%%%%%%%%%%%%%%%%%%%%%%%%%%%%%%%%%%%%%%%%%%%%%%%%%%%%%%%%%%%%%%%%
One-dimensional (1D) quantum spin systems with an energy gap above a singlet ground state have attracted a great amount of attention both theoretically and experimentally. When the magnetic field is applied, the energy gap vanishes at the lower critical field $H_{\rm c1}$ due to the Zeeman effect and a quantum phase transition from a gapped spin liquid to the Tomonaga-Luttinger liquid (TLL) takes place \cite{sakai-takahashi,yajima,sakai,chitra,usami1,usami2,kuramoto,usami3,giamarchi,furusaki}. For some typical 1D gapped spin systems, it was verified theoretically that the critical exponent of the spin correlation function in $H_{\rm c1}<H<H_{\rm c2}$ shows characteristic field dependence in each model \cite{sakai-takahashi,yajima,sakai,chitra,usami1,usami2,kuramoto,usami3,giamarchi,hikihara,konik,fath,suzuki,maeshima,suzuki1} with $H_{\rm c2}$ being the upper critical field corresponding to the saturation of the magnetization. 
Furthermore, it was argued that such characteristic field dependence of the critical exponent can be detected by NMR measurements in the TLL regime \cite{chitra,haga}. When temperature is decreased in the TLL regime, the NMR relaxation rate $1/T_1$ shows divergence behavior and its exponent is described as a function of the critical exponent of the spin correlation function. 
When the NMR measurement is performed on the nuclei located at the sites different from the electronic spins, relaxation occurs through a dipolar interaction between the nuclear and electron spins. In this case, $1/T_1$ is expressed as a sum of the longitudinal and transverse relaxation processes in magnetic fields. The divergence of $1/T_1$ with decreasing temperature is caused by one of these relaxation processes \cite{haga}.

Stimulated by these theoretical studies, NMR relaxation rates in $H_{\rm c1}<H<H_{\rm c2}$ were measured in the Haldane-gap compound ${\rm (CH_3)_4NNi(NO_2)_3}$ \cite{goto}, the $S=1/2$ bond-alternating spin-chain compound pentafluorophenyl nitronyl nitroxide (${\rm F_5PNN}$) \cite{izumi}, and the $S=1$ bond-alternating spin-chain compound Ni(C$_9$H$_{24}$N$_4$)(NO$_2$)ClO$_4$ \cite{ntenp}. 
In these experiments, the increase of $1/T_1$ in the gapless regime was observed with decreasing temperature. However, the field dependence of the divergence exponent derived from the experiments is still controversial. 
To develop precise evaluation, it is indispensable to clarify the field and temperature dependences of factors other than the divergence ones in the expression of $1/T_1$.

In this paper, we investigate the NMR relaxation rate of quantum spin chains in the TLL state. 
In Sec. II, we evaluate the field and temperature dependences of factors other than the divergence ones of $1/T_1$ in the TLL state. A criterion that the temperature dependence appears only in the power-law behavior is obtained. 
In Sec. III, we discuss the divergence behavior of $1/T_1$ in connection with theoretical results obtained so far for some 1D gapped spin systems in magnetic fields. We further discuss the experimental results in comparison with the present results.

%%%%%%%%%%%%%%%%%%%%%%%%%%%%%%%%%%%%%%%%%%%%%%%%%%%%%%%%%%%%%%%%%%%%%%%%%%%%%
\section{Field and temperature dependences of $1/T_1$ in TLL state}
%%%%%%%%%%%%%%%%%%%%%%%%%%%%%%%%%%%%%%%%%%%%%%%%%%%%%%%%%%%%%%%%%%%%%%%%%%%%%
Let us start our discussion from the formula of $1/T_1$: 
%%%%%%%%%%%%%%%%%%%%%%
\begin{eqnarray}
\frac{1}{T_1} = \int \frac{{\rm d}q}{2\pi} \lim_{\omega \rightarrow 0} 
        \frac{2k_{\rm B}T}{\hbar^2 \omega}
        \left[ F^{zz}(q) {\rm Im} \chi_{\rm R}^{zz}(q,\omega)  
                  + F^{\bot}(q) {\rm Im} \chi_{\rm R}^{\bot}(q,\omega) \right], \label{nmr}
\end{eqnarray}
%%%%%%%%%%%%%%%%%%%%%%
where $F^{zz}(q)$ and $F^{\bot}(q)$ are the longitudinal and transverse components of the hyperfine form factor, and $\chi_{\rm R}^{zz}(q,\omega)$ and $\chi_{\rm R}^{\bot}(q,\omega)$ are the longitudinal and transverse dynamical susceptibilities defined as 
%%%%%%%%%%%%%%%%%%%%%%
\begin{eqnarray}
\chi_{\rm R}^{zz}(q,\omega) = \int {\rm d}x \int {\rm d}t
     \langle S^z(x,\tau)S^z(0,0) \rangle |_{\tau \rightarrow it+0^+}  
     {\rm e}^{i(\omega t - qx)}, 
\label{chiz}
\end {eqnarray}
\begin{eqnarray}
\chi_{\rm R}^{\bot}(q,\omega) = \frac{1}{2}\int {\rm d}x \int {\rm d}t
     \left[\langle S^+(x,\tau)S^-(0,0) \rangle
   + \langle S^-(x,\tau)S^+(0,0) \rangle \right]_{\tau \rightarrow it+0^+} 
     {\rm e}^{i(\omega t - qx)}. 
\label{chix}
\end{eqnarray}
%%%%%%%%%%%%%%%%%%%%%%
Spin correlation functions in the one-component TLL are expressed as 
%%%%%%%%%%%%%%%%%%%%%%
\begin{eqnarray}
\langle S^z(x,\tau)S^z(0,0) \rangle  
   = m^2 + C_{1} \left[(x+iv\tau)^{-2}+ (x-iv\tau)^{-2} \right] 
   + C_{2}\cos(2k_{\rm F}x) (x^2+v^2\tau^2)^{-2g} \cdots,
\label{corrz}
\end{eqnarray}
\begin{eqnarray}
\lefteqn{
  \langle S^+(x,\tau)S^-(0,0) \rangle + \langle S^-(x,\tau)S^+(0,0) \rangle 
  =     C_{3}\cos(\pi x) (x^2+v^2\tau^2)^{-1/2g}  }
\nonumber \\
   & &  +  C_{4}\cos(\pi x) (x^2+v^2\tau^2)^{-1/2g-2g}
           \big[{\rm e}^{i2k_{\rm F}x} (x+iv\tau)^{-2}
            +{\rm e}^{-i2k_{\rm F}x}(x-iv\tau)^{-2} \big] \cdots, 
\label{corrx}
\end{eqnarray}
%%%%%%%%%%%%%%%%%%%%%%
where $m$ is the magnetization satisfying $0 \leq m \leq 1/2$, $2k_{\rm F}=\pi (1-2m)$, and $v$ and $\tau$ are the velocity of the spinon excitation and the imaginary time, respectively. 
$C$'s are positive constants, which have solely field dependence characteristic of the model.

The expressions for $\chi_{\rm R}^{zz}(q,\omega)$ and $\chi_{\rm R}^{\bot}(q,\omega)$ at finite temperatures were obtained in Ref. \cite{chitra} by conformally mapping $v\tau \pm ix$ onto the Matsubara strip as 
$v\tau \pm ix \rightarrow (\hbar v/\pi k_{\rm B}T) \sin [(v\tau \pm ix)\pi k_{\rm B}T/\hbar v]$ 
in (\ref{corrz}) and (\ref{corrx}), and then performing Fourier transform with respect to $x$ and $t$ in (\ref{chiz}) and (\ref{chix}). 
We assume that the $q$ dependence of the hyperfine form factor is weaker than that of the dynamical susceptibility around the gapless point and that the hyperfine form factor takes the value at the wave number of the gapless point. 
Using the dynamical susceptibilities and this assumption, we obtain the following expression for $1/T_1$, 
%%%%%%%%%%%%%%%%%%%%%%
\begin{eqnarray}
\frac{1}{T_1}  &=&
  C_{1} \left( \frac{2\pi k_{\rm B}T}{\hbar v} \right) 
  \left[ \frac{1}{2\pi^2 \hbar v} F^{zz}(0) \right] \frac{1}{4\pi}
\nonumber \\
&+& C_{2} \left( \frac{2\pi k_{\rm B}T}{\hbar v} \right)^{2g-1} 
  \left[ \frac{1}{2\pi^2 \hbar v}  F^{zz}(2k_{\rm F}) \right] 
  \lim_{\tilde{\omega} \rightarrow 0} \int {\rm d}\tilde{q} \, 
  {\rm Im} \Pi(\tilde{\omega},
           \delta \tilde{\epsilon}_{2k_{\rm F}}(\tilde{q});2g)
\nonumber \\
&+& C_{3} \left( \frac{2\pi k_{\rm B}T}{\hbar v} \right)^{1/2g-1} 
  \left[ \frac{1}{2\pi^2 \hbar v}  F^{\bot}(\pi) \right] 
  \lim_{\tilde{\omega} \rightarrow 0} \int {\rm d}\tilde{q} \, 
      {\rm Im} \Pi \left(\tilde{\omega},
               \delta \tilde{\epsilon}_{\pi}(\tilde{q});\frac{1}{2g} \right)
\nonumber \\
&+& C_{4} \left( \frac{2\pi k_{\rm B}T}{\hbar v} \right)^{2\gamma +1} 
  \left[ \frac{1}{2\pi^2 \hbar v}  
         F^{\bot}(\pi-2k_{\rm F}) \right] 
\lim_{\tilde{\omega} \rightarrow 0} \int {\rm d}\tilde{q} \, 
  {\rm Im} \Lambda(\tilde{\omega},
           \delta \tilde{\epsilon}_{\pi-2k_{\rm F}}(\tilde{q});2\gamma), 
\label{nmrf}
\end{eqnarray}
%%%%%%%%%%%%%%%%%%%%%%
where $2\gamma=2g+1/2g-2$, $\delta \tilde{\epsilon}_{k}(\tilde{q})=|\tilde{q}-\tilde{k}|$ denotes the linear dispersion relation of the spinon excitation around the gapless point $q=k \: (k=2k_{\rm F}, \pi$, and $\pi-2k_{\rm F})$, 
$\tilde{\omega}= \hbar \omega/4\pi k_{\rm B}T$ and $\tilde{q}=\hbar vq/4\pi k_{\rm B}T$ are the dimensionless variables, and
%%%%%%%%%%%%%%%%%%%%%%
\begin{eqnarray}
\Pi(\tilde{\omega},\delta \tilde{\epsilon}_k(\tilde{q});\alpha) 
 = \frac{1}{\tilde{\omega}} \sin(\frac{\pi \alpha}{2}) 
    B\left( -i\left[\tilde{\omega}+\delta \tilde{\epsilon}_k(\tilde{q})\right]
    +\frac{\alpha}{4}, 1-\frac{\alpha}{2} \right) \,
    B\left( -i\left[\tilde{\omega}-\delta \tilde{\epsilon}_k(\tilde{q})\right]
    +\frac{\alpha}{4}, 1-\frac{\alpha}{2} \right),
\label{gamma1}
\end{eqnarray}
%%%%%%%%%%%%%%%%%%%%%%
\begin{eqnarray}
\lefteqn{
\Lambda(\tilde{\omega},
        \delta \tilde{\epsilon}_{\pi-2k_{\rm F}}(\tilde{q});2\gamma) 
} \nonumber \\
 &=& \frac{1}{\tilde{\omega}} \sin(\pi \gamma) \bigg\{
     B\left( -i[\tilde{\omega}
         +\delta \tilde{\epsilon}_{\pi-2k_{\rm F}}(\tilde{q})]
         +\frac{\gamma}{2}+1, -1-\gamma \right) \,
     B\left( -i[\tilde{\omega}
         -\delta \tilde{\epsilon}_{\pi-2k_{\rm F}}(\tilde{q})]
         +\frac{\gamma}{2}, 1-\gamma \right)
\nonumber \\
 &+& B\left( -i[\tilde{\omega}
         +\delta \tilde{\epsilon}_{\pi-2k_{\rm F}}(-\tilde{q})]
         +\frac{\gamma}{2}+1, -1-\gamma \right) \,
     B\left( -i[\tilde{\omega}
         -\delta \tilde{\epsilon}_{\pi-2k_{\rm F}}(-\tilde{q})]
         +\frac{\gamma}{2}, 1-\gamma \right)
      \bigg\}, 
\label{gamma2}
\end{eqnarray}
%%%%%%%%%%%%%%%%%%%%%%
with $B(x,y)$ being the beta function. 
Note that the exponents of the power-law behavior about $T$ were obtained in Ref. \cite{chitra}.

%%%%%%%%%%%%%%%%%%%%%%%%%%%%%%%%%%%%%%
\begin{figure}[thb]
\begin{center}
\includegraphics[trim=12mm 85mm 0mm 30mm,scale =0.42]{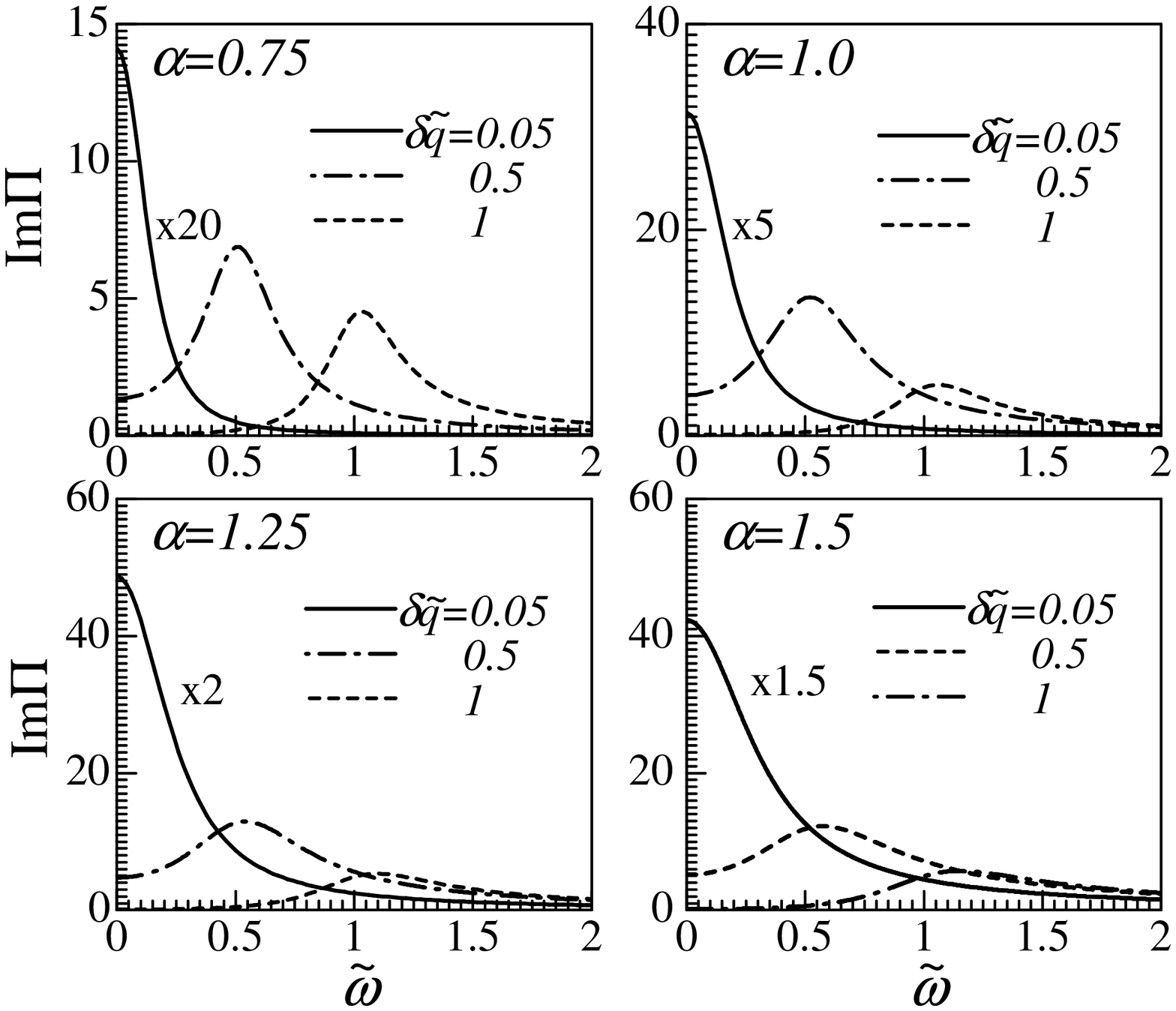}
\includegraphics[trim=12mm 85mm 0mm 30mm,scale =0.42]{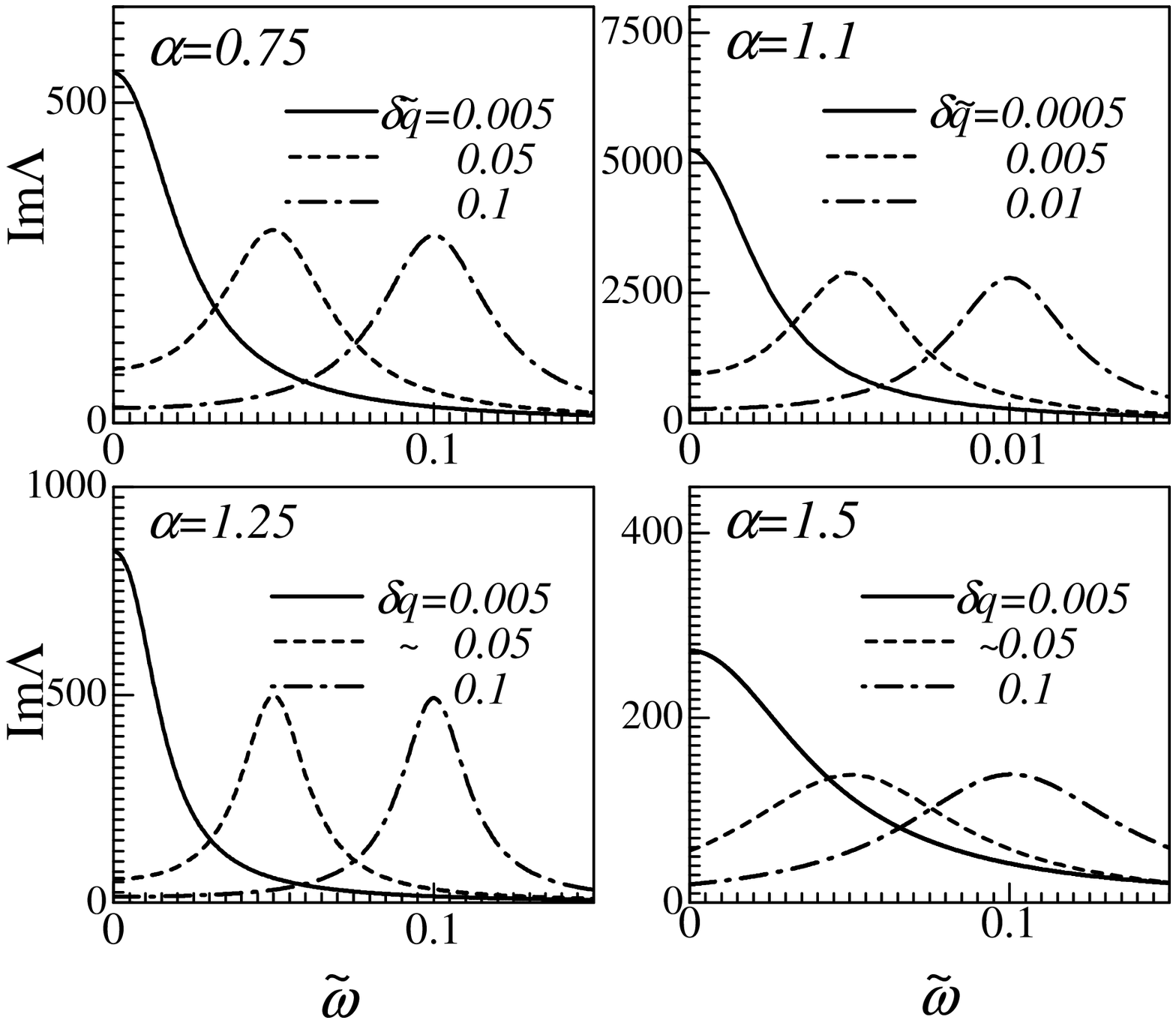}
\end{center}
\vspace{-5mm}
\caption{
The spectral weights of ${\rm Im} \Pi(\tilde{\omega},\delta \tilde{\epsilon}_k(\tilde{q});\alpha)$ and ${\rm Im} \Lambda(\tilde{\omega},\delta \tilde{\epsilon}_{\pi-2k_{\rm F}}(\tilde{q});2\gamma)$ for several values of $\alpha$, where $\delta \tilde{q}=|\tilde{q}-\tilde{k}|$ and $2\gamma=\alpha+1/\alpha-2$. 
}
\label{Fig1}
\end{figure}
%%%%%%%%%%%%%%%%%%%%%%%%%%%%%%%%%%%%%%
In Fig. 1, we show the spectral weights of ${\rm Im} \Pi(\tilde{\omega},\delta \tilde{\epsilon}_k(\tilde{q});\alpha)$ and ${\rm Im} \Lambda(\tilde{\omega},\delta \tilde{\epsilon}_{\pi-2k_{\rm F}}(\tilde{q});2\gamma)$ for several values of $\alpha$, where $2\gamma=\alpha+1/\alpha-2$. 
Around the gapless point $\delta \tilde{q}=|\tilde{q}-\tilde{k}| \sim 0$, the peak appears at $\tilde{\omega}=0$, indicating the presence of the overdamped spinon excitation. The NMR relaxation is dominated by the contribution of these excitations \cite{sachdev}. 
As $\alpha$ is decreased in ${\rm Im} \Pi(\tilde{\omega},\delta \tilde{\epsilon}_k(\tilde{q});\alpha)$ and $\alpha$ approaches unity in ${\rm Im} \Lambda(\tilde{\omega},\delta \tilde{\epsilon}_{\pi-2k_{\rm F}}(\tilde{q});2\gamma)$, the overdamped peak grows and its width becomes narrow. 
Note that for $g=1/2$ ($\alpha=1$), where the divergence of $1/T_1$ vanishes, $\Pi(\tilde{\omega},\delta \tilde{\epsilon}_{\pi}(\tilde{q});1)$ takes the same form as the universal dynamical staggered susceptibility of an SU(2) spin chain obtained in Ref. \cite{sachdev}:
$
\Pi(\tilde{\omega},\delta \tilde{\epsilon}_{\pi}(\tilde{q});1) 
 = \frac{\pi}{\tilde{\omega}} 
   \frac 
    {\Gamma \left(1/4-i\left[\tilde{\omega}
            +\delta \tilde{\epsilon}_{\pi}(\tilde{q})\right] \right)
    \Gamma \left(1/4-i\left[\tilde{\omega}
            -\delta \tilde{\epsilon}_{\pi}(\tilde{q})\right] \right)}
    {\Gamma \left(3/4-i\left[\tilde{\omega}
            +\delta \tilde{\epsilon}_{\pi}(\tilde{q})\right] \right)
    \Gamma \left(3/4-i\left[\tilde{\omega}
            -\delta \tilde{\epsilon}_{\pi}(\tilde{q})\right] \right)}.    
$
The spectral profile of its imaginary part adequately scaled \cite{comm} agrees well with each other.

%%%%%%%%%%%%%%%%%%%%%%%%%%%%%%%%%%%%%%
\begin{figure}[thb]
\begin{center}
\includegraphics[trim=2mm 185mm 0mm 10mm,scale=0.47]{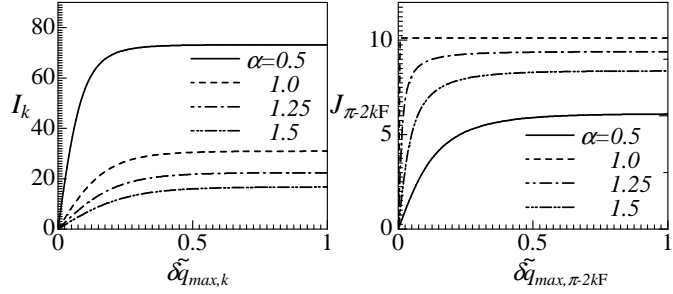}
\end{center}
\vspace{-6mm}
\caption{
The integrated values of $I_k(\alpha;T)$ and $J_{\pi-2k_{\rm F}}(2\gamma;T)$ as a function of $\delta \tilde{q}_{{\rm max},k} \equiv |\tilde{q}_{\rm max} - k|$, where $k=2k_{\rm F}, \pi$, and $\pi-2k_{\rm F}$ and $2\gamma=\alpha+1/\alpha-2$. }
\label{Fig2}
\end{figure}
%%%%%%%%%%%%%%%%%%%%%%%%%%%%%%%%%%%%%%
We next investigate the integrated values with respect to $\tilde{q}$ in the expression (\ref{nmrf}). For the appearance of the TLL state, the small $q$ region of the linear dispersion curve around the gapless point $k(=2k_{\rm F}, \pi$, and $\pi-2k_{\rm F})$ has to be considered. 
We thus introduce the upper cutoff of the $\tilde{q}$ integral around the gapless point as 
$I_k(\alpha;T)=\lim_{\tilde{\omega} \rightarrow 0} 
\int_{\tilde{k}}^{\tilde{q}_{{\rm max},k}} {\rm d}\tilde{q} \, {\rm Im} 
\Pi(\tilde{\omega},\delta \tilde{\epsilon}_{k}(\tilde{q});\alpha)$ 
and 
$J_{\pi-2k_{\rm F}}(2\gamma;T)=\lim_{\tilde{\omega} \rightarrow 0} \int_{\tilde{\pi}-2\tilde{k}_{\rm F}}^{\tilde{q}_{{\rm max},\pi-2k_{\rm F}}} {\rm d} \tilde{q} \, {\rm Im} \Lambda(\tilde{\omega},\delta \tilde{\epsilon}_{\pi-2k_{\rm F}}(\tilde{q});2\gamma)$. As temperature is decreased, $\tilde{q}_{{\rm max},k}$ becomes large. 
In Fig. 2, $I_k(\alpha;T)$ and $J_{\pi-2k_{\rm F}}(2\gamma;T)$ are shown as a function of $\delta \tilde{q}_{{\rm max},k} \equiv |\tilde{q}_{{\rm max},k} - k| (=\delta \tilde{\epsilon}_k(\tilde{q}_{{\rm max},k}))$. 
For small $\delta \tilde{q}_{{\rm max},k}$, $I_k(\alpha;T)$ and $J_{\pi-2k_{\rm F}}(2\gamma;T)$ depend on $\delta \tilde{q}_{{\rm max},k}$. The results indicate that the temperature dependence appears not only in the power-law factors but also in their coefficients. 
For $\delta \tilde{q}_{{\rm max},k} > 0.4$, on the other hand, $I_k(\alpha;T)$ and $J_{\pi-2k_{\rm F}}(2\gamma;T)$ take constants, which depend only on $\alpha$. 
Since $\alpha (=2g$  or $1/2g)$ has solely field dependence, $I_{2k_{\rm F}}(2g;T), I_{\pi}(1/2g;T)$, and $J_{\pi-2k_{\rm F}}(2\gamma;T)$ are independent of temperature. 
In this case, therefore, {\it the temperature dependence of $1/T_1$ appears solely in the power-law behavior} as 
%%%%%%%%%%%%%%%%%%%%%%
\begin{eqnarray}
\frac{1}{T_1}  &=&
  C_{1} \left( \frac{2\pi k_{\rm B}T}{\hbar v} \right) 
  \left[ \frac{1}{2\pi^2 \hbar v} F^{zz}(0) \right] \frac{1}{4\pi}
+ C_{2} \left( \frac{2\pi k_{\rm B}T}{\hbar v} \right)^{2g-1} 
  \left[ \frac{1}{2\pi^2 \hbar v}  F^{zz}(2k_{\rm F}) \right] 
  I_{2k_{\rm F}}(2g)
\nonumber \\
&+& C_{3} \left( \frac{2\pi k_{\rm B}T}{\hbar v} \right)^{1/2g-1} 
  \left[ \frac{1}{2\pi^2 \hbar v}  F^{\bot}(\pi) \right] 
  I_{\pi}\left( \frac{1}{2g} \right) 
\nonumber \\
&+& C_{4} \left( \frac{2\pi k_{\rm B}T}{\hbar v} \right)^{2\gamma +1} 
  \left[ \frac{1}{2\pi^2 \hbar v}  
         F^{\bot}(\pi-2k_{\rm F}) \right] J_{\pi-2k_{\rm F}}(2\gamma). 
\label{nmre}
\end{eqnarray}
%%%%%%%%%%%%%%%%%%%%%%
In Fig. 3, we show the $g$ dependences of $I_{2k_{\rm F}}(2g), I_{\pi}(1/2g)$, and $J_{\pi-2k_{\rm F}}(2\gamma)$. 
The corresponding factor $1/4\pi$ of the first term in (\ref{nmre}) is negligibly small in this scale. 
We find that for $g > 1/2$ the factor $(2\pi k_{\rm B}T/\hbar v)^{1/2g-1}$ in the third term in (\ref{nmre}) shows divergence and its factor $I_{\pi}(1/2g)$ becomes the largest, while for $g < 1/2$ the factor $(2\pi k_{\rm B}T/\hbar v)^{2g-1}$ in the second term shows divergence and its factor $I_{2k_{\rm F}}(2g)$ becomes the largest. 
{\it The results are universal and hold irrespective of the model}. 
Features of the model emerge in the field dependences of $g$, $v$ and $C$'s in the expression (\ref{nmre}).
%%%%%%%%%%%%%%%%%%%%%%%%%%%%%%%%%%%%%%
\begin{figure}[thp]
\begin{center}
\includegraphics[trim=25mm 150mm 25mm 0mm, scale=0.4]{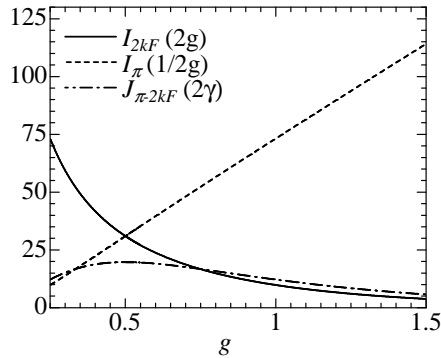}
\end{center}
\vspace{-5mm}
\caption{
The $g$ dependence of $I_{2k_{\rm F}}(2g), I_{\pi}(1/2g)$, and $J_{\pi-2k_{\rm F}}(2\gamma)$, where $2\gamma=2g+1/2g-2$. 
}
\label{Fig3}
\end{figure}
%%%%%%%%%%%%%%%%%%%%%%%%%%%%%%%%%%%%%%

If we assume that the hyperfine interaction is isotropic and $F^{zz}(0) \sim F^{zz}(2k_{\rm F}) \sim F^{\bot}(\pi) \sim F^{\bot}(\pi-2k_{\rm F})$ as usual, the factors parenthesized by $[\cdots]$ of the four terms in (\ref{nmre}) have almost the same values in a given magnetic field. 
Since the temperature dependence of $1/T_1$ is measured in a fixed magnetic field, the factors other than those concerning the power-law of $T$ in (\ref{nmre}) can be regarded as constants.

%%%%%%%%%%%%%%%%%%%%%%%%%%%%%%%%%%%%%%%%%%%%%%%%%%%%%%%%%%%%%%%%%%%%%%%%%%%%%
\section{Discussion}
%%%%%%%%%%%%%%%%%%%%%%%%%%%%%%%%%%%%%%%%%%%%%%%%%%%%%%%%%%%%%%%%%%%%%%%%%%%%%
We now discuss the divergence behavior of $1/T_1$ in connection with theoretical results obtained so far for some 1D gapped spin systems in magnetic fields. 
By making use of field theoretical and numerical techniques, the field dependence of $g$ was successfully obtained in several models. 
It was verified that $g>1/2$ is satisfied in the TLL regime of the $S=1$ isotropic \cite{sakai-takahashi,konik,fath} and anisotropic \cite{yajima} spin chains, the $S=1/2$ bond-alternating spin chain \cite{sakai,suzuki}, the $S=1/2$ two-leg spin ladder \cite{usami2,usami3,giamarchi,hikihara}, and the $S=1$ bond-alternating spin chain \cite{suzuki1}. In these models, the transverse staggered spin correlation is dominant in $H_{\rm c1}<H<H_{\rm c2}$. 
Therefore, the divergence behavior of the NMR relaxation rate is fitted well with $1/T_1 = A^{\bot}T^{1/2g-1}$.

On the other hand, in the $S=1/2$ bond-alternating spin chain with a next-nearest-neighbor interaction, the region where $g<1/2$ is satisfied emerges around the half field between $H_{\rm c1}$ and $H_{\rm c2}$ \cite{usami1,suzuki,maeshima}. In other fields, $g>1/2$ is satisfied. Such a feature is caused by the change in the dominant spin correlation in magnetic fields: Around $H \sim (H_{\rm c1}+H_{\rm c2})/2$ the longitudinal $2k_{\rm F}$ spin correlation is dominant, while  in other fields the transverse staggered spin correlation is dominant \cite{suzuki}. In this system, accordingly, the divergence behavior of $1/T_1$ around $H \sim (H_{\rm c1}+H_{\rm c2})/2$ is fitted well with $1/T_1 = A^{zz}T^{2g-1}$ or $1/T_1 = A^{\bot}T^{1/2g-1}+A^{zz}T^{2g-1}$.  
In other region of magnetic fields, in particular $H \sim H_{\rm c1}$ and $H_{\rm c2}$,  $1/T_1 = A^{\bot}T^{1/2g-1}$ is adequate.

As mentioned above, {\it a criterion for the appearance of the temperature dependence only in the power-law divergence of} $1/T_1$ is evaluated to be $\delta \tilde{q}_{{\rm max},k}>0.4$. 
We compare the temperature region derived from this criterion with that observed in the experiments. 
In the Haldane-gap compound ${\rm (CH_3)_4NNi(NO_2)_3}$, the divergence of $1/T_1$ was observed in $0.6 {\rm K}<T<2 {\rm K}$ \cite{goto}. According to the numerical calculation, the dispersion curve of the $S=1$ Heisenberg model in a small magnetization region is described well as $\epsilon(q) \sim 2J|\sin q|$. 
The coupling constant is evaluated to be $J=12 {\rm K}$ \cite{goto}. 
We estimate approximately that the linear dispersion curve may hold in the low-energy region as $0 \leq \epsilon \leq 0.2 \times 2J \sim 4.8 {\rm K}$. 
Applying $4.8 {\rm K}$ into the criterion $0.4 <\delta \tilde{q}_{{\rm max},\pi} (=\delta \tilde{\epsilon}_{\pi}(\tilde{q}_{{\rm max},\pi}))=4.8 {\rm K}/4\pi T$, we evaluate the temperature region for the divergence of $1/T_1$ as $T< 0.95 {\rm K}$.
This temperature region overlaps with that observed in the experiments. Therefore, the field dependence of the divergence exponent can be well analyzed in this compound.

In ${\rm F_5PNN}$, the divergence of $1/T_1$ was observed in $0.2 {\rm K}<T<1 {\rm K}$ \cite{izumi}. From the numerical calculation, the dispersion curve of the corresponding model is described as $\epsilon(q) \sim 0.9J|\sin q|$ in a small magnetization region. The coupling constant is evaluated as $J=5.6 {\rm K}$ \cite{takahashi}. In the same way, the temperature region for the linear dispersion curve is approximately estimated as $0 \leq \epsilon \leq 0.2 \times 0.9J \sim 1 {\rm K}$, leading to the power-law divergence region as $T< 0.2 {\rm K}$. 
This temperature region lies below that observed in the experiments. To develop more precise evaluation of the divergence exponent, measurements of $1/T_1$ in lower temperatures are necessary.

In summary, we have investigated the NMR relaxation rate of quantum spin chains in magnetic fields. The field and temperature dependences of the power-law divergence of $1/T_1$ have been evaluated in the TLL regime. On the basis of the results, experimental results for some typical gapped spin chains in magnetic fields have been discussed.  
We hope that the present analyses are useful to investigate the TLL nature of quantum spin chains in magnetic fields via the field dependence of the power-law divergence of $1/T_1$.

%%%%%%%%%%%%%%%%%%%%%%%%%%%%%%%%%%%%%%%%%%%%%%%%%%%%%%%%%%%%%%%%%%%%%%%%%%%%%
\acknowledgments
%%%%%%%%%%%%%%%%%%%%%%%%%%%%%%%%%%%%%%%%%%%%%%%%%%%%%%%%%%%%%%%%%%%%%%%%%%%%%

We would like to thank professors Y. Fujii and T. Goto for useful comments and valuable discussions.

%%%%%%%%%%%%%%%%%%%%%%%%%%%%%%%%%%%%%%%%%%%%%%%%%%%%%%%%%%%%%%%%%%%%%%%%%%%%%
%                            REFERENCES                                     %
%%%%%%%%%%%%%%%%%%%%%%%%%%%%%%%%%%%%%%%%%%%%%%%%%%%%%%%%%%%%%%%%%%%%%%%%%%%%%
% Create the reference section using BibTeX
%\bibliography{nmr}

\begin{thebibliography}{99}
%
\bibitem{sakai-takahashi} 
T. Sakai and M. Takahashi, J. Phys. Soc. Jpn. \textbf{60}, 3615 (1991); 
Phys. Rev. B \textbf{43}, 13383 (1991). 
%
\bibitem{yajima} 
M. Yajima and M. Takahashi, J. Phys. Soc. Jpn. \textbf{63}, 3634 (1994).
%
\bibitem{sakai} 
T. Sakai, J. Phys. Soc. Jpn. \textbf{64}, 251 (1995). 
%
\bibitem{chitra} 
R. Chitra and T. Giamarchi, Phys. Rev. B \textbf{55}, 5816 (1997); 
T. Giamarchi, {\it Quantum Physics in One Dimension} (Oxford University Press, 2004).
%
\bibitem{usami1} 
M. Usami and S. Suga, Phys. Lett. A \textbf{240}, 85 (1998).
%
\bibitem{usami2} 
M. Usami and S. Suga, Phys. Rev. B \textbf{58}, 14401 (1998).
%
\bibitem{kuramoto} 
T. Kuramoto, J. Phys. Soc. Jpn. \textbf{67}, 1762 (1998). 
%
\bibitem{usami3} 
M. Usami and S. Suga, Phys. Lett. A \textbf{259}, 53 (1999).
%
\bibitem{giamarchi} 
T. Giamarchi and A. M. Tsvelik,: Phys. Rev. B \textbf{59}, 11398 (1999). 
%
\bibitem{furusaki} 
A. Furusaki and S.-C. Zhang, Phys. Rev. B \textbf{60}, 1175 (1999). 
%
\bibitem{hikihara} 
T. Hikihara and A. Furusaki, Phys. Rev. B \textbf{63}, 134438 (2001). 
%
\bibitem{konik} 
R. M. Konik and P. Fendley, Phys. Rev. B \textbf{66}, 144416 (2002). 
%
\bibitem{fath} 
G. F\'{a}th, Phys. Rev. B \textbf{68}, 134445 (2003). 
%
\bibitem{suzuki} 
T. Suzuki and S. Suga, Phys. Rev. B \textbf{70}, 054419 (2004). 
%
\bibitem{maeshima} 
N. Maeshima, K. Okunishi, K. Okamoto, and T. Sakai, Phys. Rev. Lett. \textbf{93}, 127203 (2004); N. Maeshima, K. Okunishi, K. Okamoto, T. Sakai, and K. Yonemitsu, J. Phys., Condens. Matter \textbf{18}, 4819 (2006). 
%
\bibitem{suzuki1} 
T. Suzuki and S. Suga, unpublished. 
%
\bibitem{haga} 
N. Haga and S. Suga, J. Phys. Soc. Jpn. \textbf{69}, 2431 (2000). 
%
\bibitem{goto} 
T. Goto, Y. Fujii, Y. Shimaoka, T. Maekawa and J. Arai, Physica B \textbf{284-288}, 1611 (2000); T. Goto, T. Ishikawa, Y. Shimaoka and Y. Fujii, Phys. Rev. B \textbf{73}, 214406 (2006). 
%
\bibitem{izumi} 
K. Izumi, T. Goto, Y. Hosokoshi and J.-P. Boucher, Physica B \textbf{329-333}, 1191 (2003). 
%
\bibitem{ntenp} 
S. Matsubara, K. Kodama, M. Takigawa and M. Hagiwara, J. Phys. Soc. Jpn. \textbf{74}, 2417 (2005). 
%
%\bibitem{sasaki} 
%T. Sasaki, Y. Fujii, H. Kikuchi, M. Chiba, Y. Yamamoto, and H. Hori, J. Mag. Mag. Mag. (2007) in press. 
%
\bibitem{sachdev} 
S. Sachdev, T. Senthil and R. Shankar, Phys. Rev. B \textbf{50}, 258 (1994);
S. Sachdev, Phys. Rev. B \textbf{50}, 13006 (1994). 
%
\bibitem{comm} 
In this study the dimensionless variables are scaled by $4\pi k_{\rm B}T$, while in Ref. \cite{sachdev} they are scaled by $k_{\rm B}T$. 
The spectral weight of ${\rm Im} \Pi(\tilde{\omega},\delta \tilde{\epsilon}_{\pi}(\tilde{q});1)$ is larger by $8\pi^2$ than `the universal spectral weight' in Ref. \cite{sachdev}.
%
\bibitem{takahashi}
M. Takahashi, Y. Hosokoshi, H. Nakano, T. Goto, M. Takahashi and M. Kinoshita, Mol. Cryst. Liq. Cryst. \textbf{306}, 111 (1997). 
%
%
%\end{references}
\end{thebibliography}
%%%%%%%%%%%%%%%%%%%%%%%%%%%%%%%%%%%%%%%%%%%%%%%%%%%%%%%%%%%%%%%%%%%%%%%%%%%%%
%\begin{references}

%

\end{document}